\documentclass{article}
\usepackage{ijcai26}

\usepackage{times}
\usepackage{soul}
\usepackage{url}
\usepackage[hidelinks]{hyperref}
\usepackage[utf8]{inputenc}
\usepackage[small]{caption}
\usepackage{graphicx}
\usepackage{amsthm}
\usepackage{booktabs}
\usepackage{algorithm}
\usepackage{algorithmic}
\usepackage[switch]{lineno}
\usepackage{tabularx}
\usepackage{booktabs}
\usepackage{multirow}
\usepackage{amsmath,amssymb}
\usepackage{xcolor}
\usepackage[utf8]{inputenc}
\usepackage{booktabs}   
\usepackage{tabularx}  
\usepackage{caption}    
\usepackage{amsmath}    
\usepackage{color, colortbl} 

\definecolor{tableheader}{gray}{0.95}
\usepackage[utf8]{inputenc}
\usepackage{booktabs}   
\usepackage{tabularx}   
\usepackage{multirow}   
\usepackage{makecell}    
\usepackage[table]{xcolor} 
\usepackage{caption}     
\usepackage[most]{tcolorbox}
\usepackage{xcolor}

\definecolor{lightpurple}{RGB}{245, 240, 255}
\definecolor{lightblue}{RGB}{201,216,242}

\newtcolorbox{purplebox}{
    enhanced,
    colback=lightblue,
    colframe=lightblue, 
    arc=2mm,      
    boxrule=0pt,          
    left=3mm,
    right=3mm,
    top=3mm,
    bottom=3mm,
    before skip=10pt,
    after skip=10pt
}
\definecolor{tablegray}{gray}{0.95}

\title{Game-Theoretic Lens on LLM-based Multi-Agent Systems}

\author{
Jianing Hao$^1$
\and
Han Ding$^2$
\and
Yuanjian Xu$^1$
\and
Tianze Sun$^3$
\and
Ran Chen$^4$
\and
Wanbo Zhang$^5$
\and
Guang Zhang$^{1,}$\thanks{Corresponding author.}
\and
Siguang Li$^{1,}$\footnotemark[1]\\
\affiliations
$^1$The Hong Kong University of Science and Technology (Guangzhou),
$^2$Beihang University\\
$^3$Harbin Institute of Technology,
$^4$OpenCSG,
$^5$Fudan University,
\emails
jhao768@connect.hkust-gz.edu.cn,
handing@buaa.edu.cn,
yxu085@connect.hkust-gz.edu.cn,
suntianze@stu.hit.edu.cn,
schen@opencsg.com,
23302010062@m.fudan.edu.cn,
guangzhang@hkust-gz.edu.cn,
siguangli@hkust-gz.edu.cn
}

\begin{document}

\maketitle

\begin{abstract}
Large language models (LLMs) have demonstrated strong reasoning, planning, and communication abilities, enabling them to operate as autonomous agents in open environments.
While single-agent systems remain limited in adaptability and coordination, recent progress has shifted attention toward multi-agent systems (MAS) composed of interacting LLMs that pursue cooperative, competitive, or mixed objectives.
This emerging paradigm provides a powerful testbed for studying social dynamics and strategic behaviors among intelligent agents.
However, current research remains fragmented and lacks a unifying theoretical foundation.
To address this gap, we present a comprehensive survey of LLM-based multi-agent systems through a game-theoretic lens.
By organizing existing studies around the four key elements of game theory: players, strategies, payoffs, and information, we establish a systematic framework for understanding, comparing, and guiding future research on the design and analysis of LLM-based MAS.
\end{abstract}

\section{Introduction}
Multi-agent systems (MAS) have long served as a foundational framework for investigating interaction, decision-making, and strategic behavior among autonomous entities.
Classical research has examined these systems through game-theoretic models~\cite{nash1950equilibrium}, distributed optimization~\cite{sandholm1999distributed}, reinforcement learning~\cite{littman1994markov}, and agent-based modeling~\cite{epstein1999agent}, providing formal tools for understanding coordination, competition, negotiation, and emergent collective behaviors.

The recent advances in large language models (LLMs) have given rise to a new paradigm of multi-agent systems, in which agents can reason~\cite{wei2022chain}, plan~\cite{huang2022language}, and communicate~\cite{li2023camel} using natural language.
In such systems, multiple LLM-based agents interact in complex and open-ended environments, exhibiting a wide spectrum of behaviors that may include cooperation~\cite{chen2023agentverse}, competition~\cite{multiagentbench_2025}, negotiation~\cite{GAMELLM_2024}, social learning, and other emergent phenomena~\cite{ashery2025emergent}.

Continuing a line of research in strategic decision-making systems~\cite{silver2016alphago}, prior studies have explored LLM-based multi-agent systems in diverse contexts, including task decomposition and planning~\cite{hong2023metagpt}, decision-making under uncertainty~\cite{agashe2025llmcoordinationevaluatinganalyzingmultiagent}, multi-agent coordination~\cite{wu2023autogen}, social simulation and behavioral modeling~\cite{park2023generative}, multi-party debate or dialogue~\cite{qian2024chatdev}, and competitive strategy games~\cite{multiagentbench_2025}.
These works highlight the potential of LM-MAS to provide insights into agent interaction, strategic reasoning, communication protocols, and emergent collective behaviors, yet a unified theoretical perspective that systematically organizes these developments is still lacking.

Behind every interaction in a MAS lies a strategic decision-making problem, which is precisely the domain that game theory is designed to model.
As LLM-powered agents become more capable and autonomous, understanding their behavior through a game-theoretic lens becomes essential.
This survey adopts a \textbf{game-theoretic lens} to systematically synthesize existing research on LLM-driven multi-agent systems.
This paper makes the following three contributions:
\begin{itemize}
    \item We introduce a novel game-theoretic framework for categorizing LLM-based MAS through four core elements: players, strategies, payoffs, and information. This unified perspective facilitates the integration of classical game theory with modern LLM-driven research for systemic analysis of agent interactions.
    \item We provide a systematic review of the current body of work on LLM-based multi-agent systems, revealing the insights and limitations of current research.
    \item We identify key research gaps and propose forward-looking research directions, focusing on optimizing equilibrium coordination, designing incentive-compatible communication protocols, and information structure modeling under partial observability.
\end{itemize}
\section{Background and Taxonomy}

\begin{figure}
    \centering
    \includegraphics[width=0.85\linewidth]{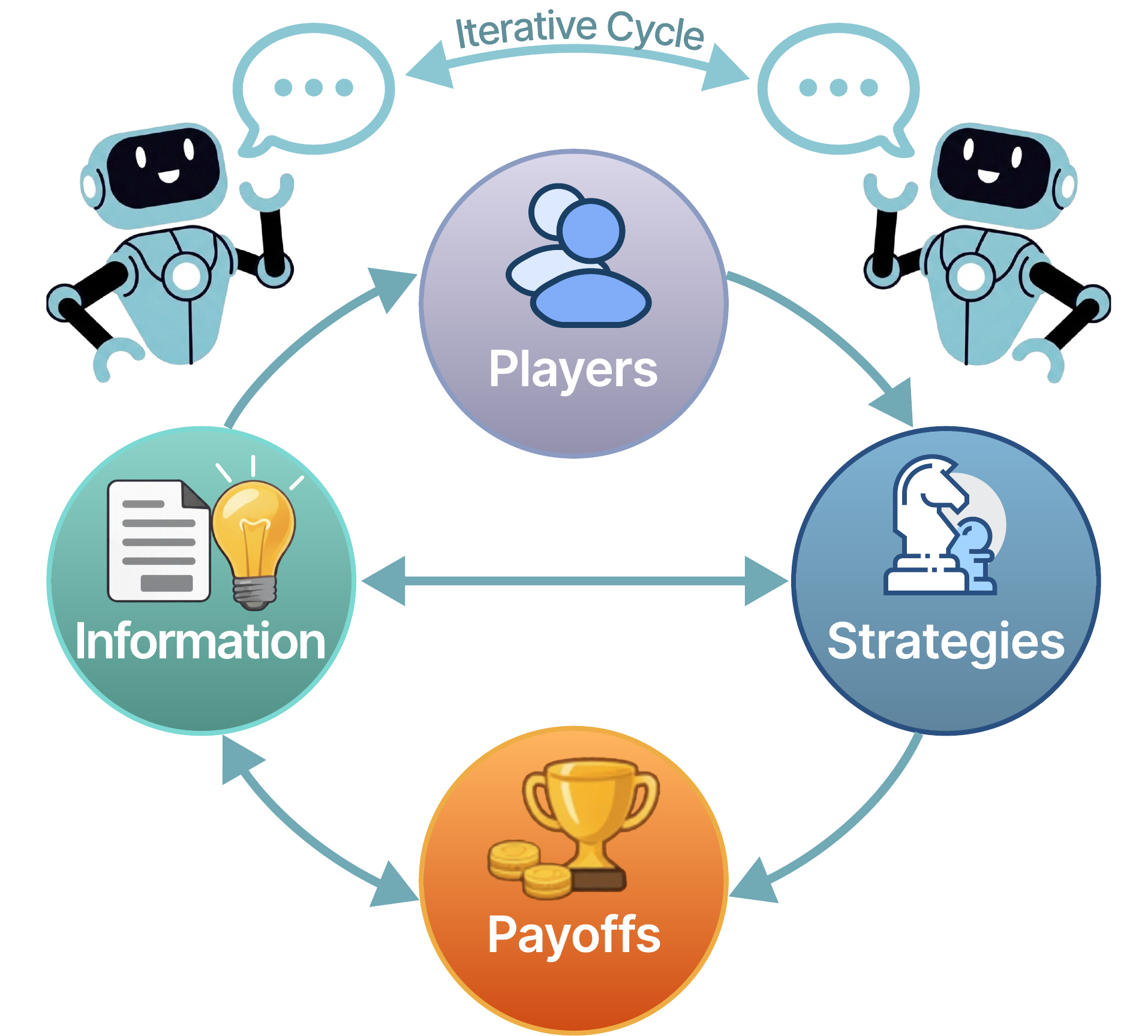}
    \vspace{-1mm}
    \caption{A game-theoretic framework for LLM-based multi-agent systems illustrates the dynamic interplay between the four core elements of a game: Players, Strategies, Payoffs, and Information.}
    \vspace{-4mm}
    \label{fig:taxo}
\end{figure}

\subsection{Foundations of Game Theory}
Game theory is the mathematical study of strategic decision-making, where multiple agents interact, and each agent's outcome depends on the choices made by others.
It studies the strategic interactions of rational decision-makers, where each player seeks to maximize their own payoff.
A normal-form game is defined by the tuple $\Gamma=(N,(S_i)_{i\in N},(u_i)_{i\in N})$, where $N=\{1,\dots,n\}$ is the finite set of players, $S_i$ is the finite set of pure strategies available to player $i$, and $u_i$:$\prod_{j\in N}S_j\to\mathbb{R}$ is player $i$'s payoff function assigning a real-valued reward to every joint strategy profile.  
A strategy profile is a vector $\mathbf{s}=(s_1,\dots,s_n)\in S_1\times\cdots S_j \cdots\times S_n$, representing the strategy choices of all players.
In this context, we assume that players are rational and aim to maximize their expected utility by selecting strategies that optimize their payoff, based on the information available to them.

Strategies may be pure or mixed.  A pure strategy for player $i$ is an element $s_i\in S_i$, while a mixed strategy is a probability distribution $\sigma_i\in\Delta(S_i)$ over $S_i$.
Here $\Delta(S_i)$ denotes the simplex of probability measures on $S_i$.
A mixed-strategy profile $\sigma=(\sigma_1,\dots,\sigma_n)$ specifies a distribution over each player's actions.
Under profile $\sigma$, each pure profile $s=(s_1,\dots,s_n)$ is chosen with probability $\prod_{j=1}^n\sigma_j(s_j)$, and the expected payoff to player $i$ is
$U_i(\sigma)\;=\;\sum_{s\in S}u_i(s)\prod_{j=1}^n\sigma_j(s_j)$.
A pure strategy $s_i$ can be viewed as a special case of a mixed strategy where the probability is concentrated on a single action.

Strategy selection is influenced by both payoffs and information.
Under complete information, all payoff functions and strategy spaces $S_i$ are common knowledge, allowing players to perfectly anticipate others' behavior.
In contrast, under incomplete information, each player may hold private knowledge or beliefs, modeled as $\theta_i$.
Regardless of the information structure, players choose strategies that maximize their expected utility based on the information available to them.
The attractiveness of strategies is determined by their associated payoffs, while the structure of available information dictates how strategies are conditioned and how beliefs are updated.
A Nash Equilibrium occurs when no player can unilaterally improve their outcome by changing strategies — assuming others keep theirs fixed. This concept is critical in MAS, where agents seek stable strategies in competitive or cooperative settings.

\subsection{Multi-Agent Systems as Game-Theoretic Environments}
A MAS consists of multiple autonomous agents interacting within a shared environment.
Unlike single-agent systems, MAS leverage distributed intelligence, allowing agents to divide work, respond to local signals, and solve complex problems through either collaboration or competition.
Traditional MAS relied on rule-based logic and local policies,  though challenges remain where local metrics may misrepresent global behavior, as similar concerns arise in learning-in-games dynamics~\cite{fudenberg1998theory}.
However, with the advent of LLMs, agents can engage in natural language reasoning, flexible task negotiation, and strategic coordination.
In the game-theoretic perspective, each agent is modeled as a player $i\in N$ with its own strategy space and utility function.
Shoham and Leyton-Brown~\cite{multiagent_2008} describe a MAS as a system of autonomous entities with potentially divergent interests or information, which mirrors the game-theoretic notion of players having different payoffs or private knowledge.
Thus, any MAS interaction can be captured as a strategic-form game: an agent's possible actions correspond to the player's strategies, and the agent's objectives are encoded by its payoff function over joint action profiles.

Communication and coordination are key features in MAS.
Communication can be modeled as pre-play signaling: agents exchange messages or share observations before choosing their actions.
Such signals allow agents to share private information or align expectations.
Coordination problems arise when agents share common goals: if their payoffs are aligned, the MAS essentially operates as a cooperative game.
More generally, coordinating self-interested agents involves finding equilibria in games where multiple consistent strategy profiles exist, or addressing mechanism-design problems that align individual incentives with collective objectives, particularly in settings where communication may carry strategic intent or misrepresentation~\cite{kamenica2011bayesian}.
Autonomy in MAS refers to the absence of a central controller. 
Each agent decides independently based on its own payoff and available information.
This reflects the non-cooperative game assumption of decentralized decision-making.
Agents may have private or incomplete information about the environment or other players' preferences, analogous to players' types in Bayesian games.
In summary, the features of MAS directly map onto game-theoretic concepts: communication equates to signaling or messaging in games, coordination mirrors signaling in games, coordination corresponds to equilibrium selection or coalition formation, and autonomy reflects independent, utility-maximizing players.

LLM-based MAS introduces unique characteristics.
In these systems, agents are driven by large language models that interact through natural language.
The effective action space consists of potential utterances or text sequences, which is combinatorial and open-ended.
Communication is inherently language-driven, enabling rich negotiation and explanation.
Recent studies have shown that LLM agents can develop communication protocols and roles spontaneously.
Through dialogue, these agents can negotiate resources, delegate tasks, or balance cooperation and competition, demonstrating emergent behaviors such as cooperation, competition, negotiation, and social interaction. This capacity for strategic communication under incentives explicitly links LLM interactions to classical signaling theory.
These characteristics suggest that LLM-based agents operate as game-theoretic players with highly expressive, language-driven strategies and dynamic information exchange, leading to more complex and dynamic interactions than traditional MAS models.

\subsection{Taxonomy based on Game-Theoretic Elements}
In MAS, agents don’t operate in isolation; they interact, influence, and respond to one another.
Game theory provides a formal framework for modeling these interactions by treating agents as strategic players whose choices are interdependent.
We organize our analysis around four core game-theoretic elements: Player, Strategy, Payoff, and Information.
These elements form the foundational components of any strategic model.
We focus on these four dimensions because together they define who the agents are, what actions they can take, what they value, and what they know.
Notably, we do not treat Action as separate from Strategy, as an action is simply a specific strategy choice.
Additionally, we exclude Equilibrium as a primitive element since it is a derived outcome based on the other four components.
In a normal-form game, the strategic interaction is fully determined by the players, their strategy sets, and their payoff functions.
We include Information as a distinct dimension to account for the agents' private knowledge or type structure, distinguishing between complete- and incomplete-information settings.

These elements interact dynamically. 
Each player $i$ has a strategy space $S_i$ and payoff $u_i$, but which strategies are chosen depends on the information available to $i$.
Payoffs dictate incentives: given the available information, a player selects the strategy that maximizes its utility.
Conversely, the strategies chosen and their outcomes can reveal additional information, thereby influencing subsequent payoffs and decisions.
Figure~\ref{fig:taxo} illustrates these interdependencies: players (P) choose strategies (S) based on payoff incentives and information signals, and the resulting profile produces payoffs and potentially new information.

\section{Game-Theoretic Elements in MAS}

\subsection{Players}
\begin{figure}
    \centering
    \includegraphics[width=0.9\linewidth]{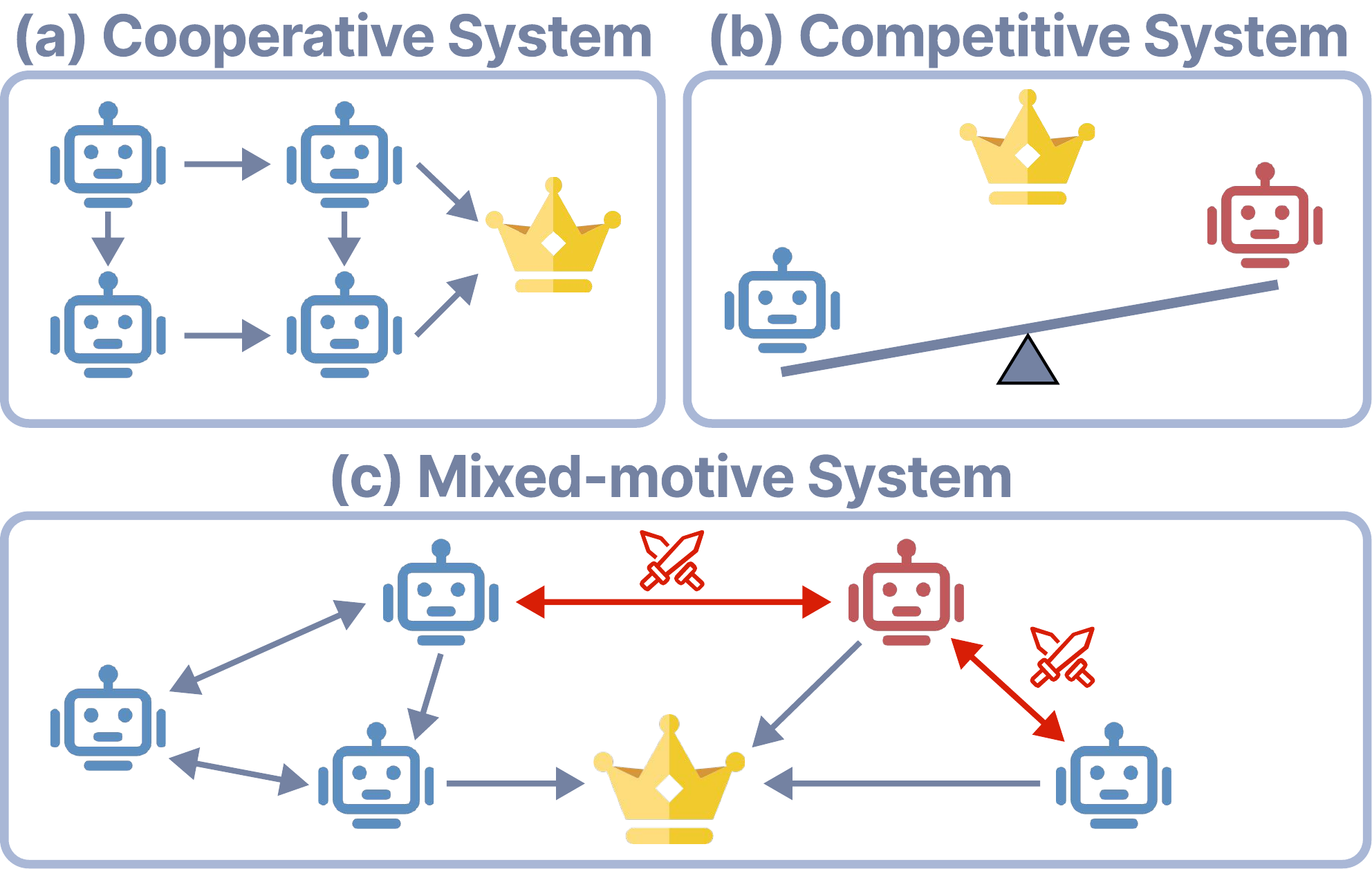}
    \vspace{-1mm}
    \caption{Illustration of three interaction structures among LLM-based players.}
    \label{fig:player}
    \vspace{-4mm}
\end{figure}

In LLM-based multi-agent systems, agents (players) engage in interactions that extend far beyond traditional two-player games.
Depending on their design and objectives, these agents may differ in capabilities, goals, or access to information, yielding heterogeneous populations that give rise to diverse strategic dynamics.
From a game-theoretic perspective, these interactions can be broadly categorized into cooperative, competitive, and mixed-motive systems, reflecting varying degrees of alignment among agents' objectives, as illustrated in Figure~\ref{fig:player}.

\paragraph{Cooperative Systems.}
Cooperative systems involve agents pursuing a shared or aligned objective, often formalized through a welfare function $W(a)$ that aggregates individual utilities.
\[
W(a) \;=\; f\big(u_1(a),\dots,u_N(a)\big),
\quad
a^* \;=\; \arg\max_{a\in A} W(a),
\]
where $A=\prod_i A_i$ denotes joint action space, and $f:\mathbb{R}^N\to\mathbb{R}$ is an aggregation map.
A common choice is a weighted sum $W(a)=\sum_i w_i u_i(a)$ (linear $f$), while complex scenarios may require nonlinear aggregation for fairness or risk-sensitive goals.
Cooperative LLM agents effectively form an optimistic team, sharing information and coordinating to find the action profile $a^*$ that maximizes joint welfare.

Recent studies have demonstrated the effectiveness of cooperative LLM agents on challenging tasks~\cite{multiagentbench_2025}.
For instance, the Chain-of-Agents (CoA) framework~\cite{zhang2024chain} decomposes a long-context problem into segments handled by sequential ``worker" agents followed by a ``manager" agent, enabling complex reasoning over extended inputs.
The COPPER~\cite{bo2024reflective} system uses self-reflective agents whose outputs are fine-tuned through counterfactual rewards to improve contribution quality.
Role-based dialogue frameworks like CAMEL~\cite{li2023camel} use ``assistant–user" role-playing to simulate human-like teamwork, generating emergent collaboration behaviors.
Other architectures, like MetaGPT~\cite{hong2023metagpt}, assign specialized roles such as project manager, architect, and developer to agents within software engineering tasks, while AutoGen~\cite{wu2023autogen} provides a flexible framework for orchestrating multi-agent interactions.
Empirical evaluations have shown that language-mediated coordination can significantly improve performance on tasks like question answering, coding, and planning.

\paragraph{Competitive Systems.}
In competitive systems, agents have conflicting objectives.
These can be formalized as (possibly zero-sum) games $G=(\mathcal{N},{A_i},{u_i})$, where each agent $i$ chooses a strategy $\sigma_i\in\Delta(A_i)$ to maximize its expected payoff $U_i(\sigma)$.
A Nash equilibrium $(\sigma_1^*,\dots,\sigma_N^*)$ satisfies the mutual best-response conditions:
\[
U_i(\sigma_i^*,\sigma_{-i}^*) \ge U_i(\sigma_i,\sigma_{-i}^*) \quad \forall i,\sigma_i.
\]
For two-player zero-sum games, this reduces to the classical minimax equilibrium:
\[
\max_{\sigma_1}\min_{\sigma_2} U_1(\sigma_1,\sigma_2)
= \min_{\sigma_2}\max_{\sigma_1} U_1(\sigma_1,\sigma_2).
\]

LLM agents have been evaluated in adversarial benchmarks and games.
For example, GTBench~\cite{duan2024gtbench} is a game-theoretic benchmark with ten classic games (board, card, negotiation, auctions) probing pure strategic reasoning.
Another testbed, GameBench~\cite{gamebench_2024}, covers nine diverse game environments, revealing that GPT-4 still struggles to match human-level performance in zero-sum games.
Specialized competitive frameworks, such as multi-agent debate systems~\cite{selfplaying_2024}, facilitate agents arguing opposing viewpoints, thereby refining their answers.
Empirical studies report that, though limitations in multi-step strategy anticipation remain, such competitive exercises help LLMs exhibit long-term planning and opponent modeling.

\paragraph{Mixed-Motive Systems.}
Most realistic multi-agent environments blend cooperation and competition.
In mixed-motive games, each agent balances self-interest and collective welfare, captured by:
\[
u_i(a) = \alpha_i\, v_i^{\mathrm{self}}(a) + (1-\alpha_i)\, v^{\mathrm{col}}(a),
\quad \alpha_i\in[0,1],
\]
where $v_i^{\mathrm{self}}$ is agent $i$’s private payoff and $v^{\mathrm{col}}$ is a shared team payoff. The parameter $\alpha_i$ encodes the degree of self-interest versus collaboration. Designing incentives in such systems often involves mechanism design: for example, introducing transfers $t_i(a)$ that reward or penalize agents so that truthful, team-oriented behavior is an equilibrium. A common requirement is incentive compatibility (IC) and budget balance (BB), consistent with classical mechanism design principles~\cite{mas1995microeconomic}:
\[
\text{IC: }\sigma_i^*\!\in\!\arg\max_{\sigma_i}\mathbb{E}[u_i(a)+t_i(a)], 
\quad 
\text{BB: }\sum_i \mathbb{E}[t_i(a)] = 0.
\]
These constraints ensure that following the intended strategy is optimal for each agent without external subsidies.

Recent work has begun to explore mixed-motive LLM systems.
Orner et al.~\cite{tang2024explaining} investigate explanation methods for agents in mixed-motive games such as variants of Diplomacy and iterated Prisoner's Dilemma, highlighting how interleaved cooperation and competition create complex social dilemmas.
Complementing these empirical investigations, Duetting et al.~\cite{mechanism_design_2024} adopt a mechanism-design perspective, framing LLMs as bidding agents within an auction for generated content that is formally designed to ensure incentive compatibility and truthful reporting.
Furthermore, the MAC-SPGG framework~\cite{everyone_2025} addresses the strategic alignment of motives through sequential public-goods games.
This framework proves that by precisely tuning public-goods rewards, a Subgame Perfect Nash Equilibrium (SPNE) can be induced to foster universal cooperation.
Collectively, these diverse methodologies—ranging from auction theory to sequential game modeling—demonstrate that carefully structured incentives can effectively align LLM motives and sustain stable collaboration, even when individual objectives diverge.

\subsection{Strategies and Equilibrium}
Effective strategy development is crucial for multi-agent LLM systems, where agents often need to account for both cooperation and competition.
Building on language-mediated strategic agents like CICERO~\cite{meta2022human}, the LLM-Nash framework~\cite{reasoning_nash_2025} models each agent as selecting reasoning prompts as strategies, defining a ``reasoning equilibrium" over the prompt space.
Unlike classical games with fully rational players, LLM–Nash captures bounded rationality by explicitly modeling the reasoning process.
It shows that these reasoning equilibria can diverge from classical Nash outcomes, reflecting the unique strategic behaviors of LLM agents.
This highlights that LLMs can converge to new equilibrium-like behaviors through iterative dialogue and prompting, effectively approximating game-theoretic reasoning in non-cooperative settings.

Self-play is another powerful tool for strategy formation.
In self-play, an agent competes against copies of itself, refining its strategy through trial and error~\cite{selfplay_survey_2024}. 
This approach has been used in multi-agent reinforcement learning (MARL), where agents learn not only by interacting with a fixed environment but also by improving their strategies over time based on their own behavior.
Recent studies show that self-play can significantly enhance LLM strategies by enabling agents to explore a wide range of tactics and counter-strategies.  Through reinforcement learning (RL) and policy optimization, LLMs can adjust their behavior to improve their outcomes in repeated or evolving strategic environments.  This iterative process allows LLMs to converge toward optimal strategies.
SPIRAL~\cite{spiral_2025} leverages this idea by having two LLM agents play multi-turn, zero-sum language games against each other. It employs a fully online, multi-turn, multi-agent RL system so that all model parameters are continually updated during self-play.
SPIRAL demonstrates that competitive self-play not only drives agents toward strong gameplay strategies (akin to reaching equilibrium) but also endows them with general problem-solving tactics.
MARSHAL~\cite{MARSHAL_2025} framework similarly shows that self-play in strategic games can generalize to diverse reasoning tasks. 
MARSHAL uses turn-level advantage estimation and self-play in both cooperative and competitive games, finding that it strengthens LLM strategic abilities and yields gains on multi-agent benchmarks.

\subsection{Payoffs}

In multi-agent systems, the payoff structure plays a critical role in designing effective mechanisms.
Each agent’s goal is to maximize its own reward through some strategies, but in a multi-agent setting, the rewards of the agents are often not aligned.
The challenge lies in designing appropriate reward mechanisms that align individual objectives with the collective goals, ensuring that agents' behavior is optimized for both individual success and system-wide efficiency.

\paragraph{Mechanism Design.}
Effective mechanism design requires that the incentive structures of agents align with the system's collective objective.
In a multi-agent environment, let each agent $i$ have an individual reward function $R_i(s,a)$, and the system's overall objective is represented by a social welfare function $W(s,a) = f\big(R_1(s,a), \dots, R_N(s,a)\big)$, where $f(\cdot)$ might represent a combination of fairness, efficiency, or safety considerations, aligning conceptually with welfare maximization in microeconomics~\cite{mas1995microeconomic}.
Misalignment arises when maximizing $R_i$ does not necessarily improve $W$, and this misalignment can lead to suboptimal outcomes for the system.

To quantify this misalignment, we define each agent’s deviation gain under a joint policy $\pi = (\pi_1, \dots, \pi_N)$ as
\[
\Delta_i(\pi) = 
\max_{\pi_i'} 
\mathbb{E}\!\left[ R_i(s,a_i',a_{-i}) \right]
- 
\mathbb{E}\!\left[ R_i(s,a) \right],
\]
which indicates how much agent $i$ can gain by unilaterally deviating from the joint policy.
A cooperative mechanism should minimize these deviation gains while improving the collective objective. The general alignment framework can be expressed as:
\[
\begin{aligned}
\max_{\mathcal{M}} \quad & 
\mathbb{E}\!\left[ W(s,a) \right] \\
\text{s.t.} \quad &
\mathbb{E}\!\left[ R_i(s,a) \right] 
\ge 
\mathbb{E}\!\left[ R_i(s,a_i',a_{-i}) \right], 
\quad \forall i,\; \forall \pi_i',
\end{aligned}
\]
where $\mathcal{M}$ represents the designed reward or transfer mechanism ensuring incentive compatibility.
This alignment mechanism ensures that agents' behaviors are motivated by the collective good while minimizing selfish deviations.

\paragraph{Reward Shaping.}
Reward shaping provides a practical way to adjust agent incentives while preserving local autonomy.
A shaped reward might take the form:
\[
R_i'(s,a,s') = R_i(s,a,s') + \gamma\,\Phi_i(s') - \Phi_i(s),
\]
where $\Phi_i(s)$ is a potential function that adjusts the temporal credit assignment without altering the cooperative optimum.
Through this approach, reward shaping can accelerate convergence and maintain stability across agents with heterogeneous goals.
For instance, COPPER~\cite{bo2024reflective} introduces self-reflection as a method for improving cooperation, where agents adjust their incentives internally to better fulfill collaborative tasks.
While no explicit mathematical form of reward adjustment is given, this approach can be understood as the introduction of new incentive mechanisms that encourage agents to reflect on their contributions to cooperation.
In combination with mechanism design, reward shaping ensures that agents remain aligned with system-level goals while maintaining their individual autonomy. This method is particularly useful in multi-agent systems where agents must balance personal incentives with collective performance.

\paragraph{Penalty and Regulation Mechanisms.}
In addition to reward shaping, penalty and regulation mechanisms are essential tools for regulating agent behavior.\
By introducing a penalty function $\Psi_i(s,a)$, undesirable behaviors such as ethical violations or resource misuse can be discouraged, further driving agents toward socially beneficial behaviors. The regulated reward can then be expressed as:
\[
\tilde{R}_i(s,a,s') = R_i'(s,a,s') - \lambda_i\,\Psi_i(s,a),
\]
where $\lambda_i > 0$ controls the strength of the regulatory enforcement. This formulation integrates soft constraints directly into the learning signal, allowing agents to trade off performance and compliance dynamically.
For safety-critical or ethically sensitive systems, such as those involving human interactions or resource management, penalty-based regulation is essential to ensure that agents do not exploit or harm the system.
Explicit regulatory constraints can also be imposed at the policy level, where actions violating safety or ethical norms are excluded from the feasible set:
\[
\pi_i(a|s) = 0 \quad \text{if} \quad \Psi_i(s,a) > \tau_i,
\]
where $\tau_i$ is a threshold above which actions are deemed unacceptable.

Recent studies~\cite{gradient2024risk,MARL_2024,shen2024proagent} have highlighted the importance of incorporating regulatory signals in multi-agent systems, showing that integrating explicit or implicit regulatory feedback—whether linguistic, feedback-based, or self-penalization—can substantially improve both ethical compliance and long-term system stability.
Empirical frameworks such as FinCon~\cite{zhang2024fincon}, CORY~\cite{hu2024cory}, and ProAgent~\cite{shen2024proagent} demonstrate how structured reward signals, verbal feedback, or reflective evaluation can operationalize such alignment strategies in practice.

\subsection{Information}

\begin{figure}
    \centering
    \includegraphics[width=0.8\linewidth]{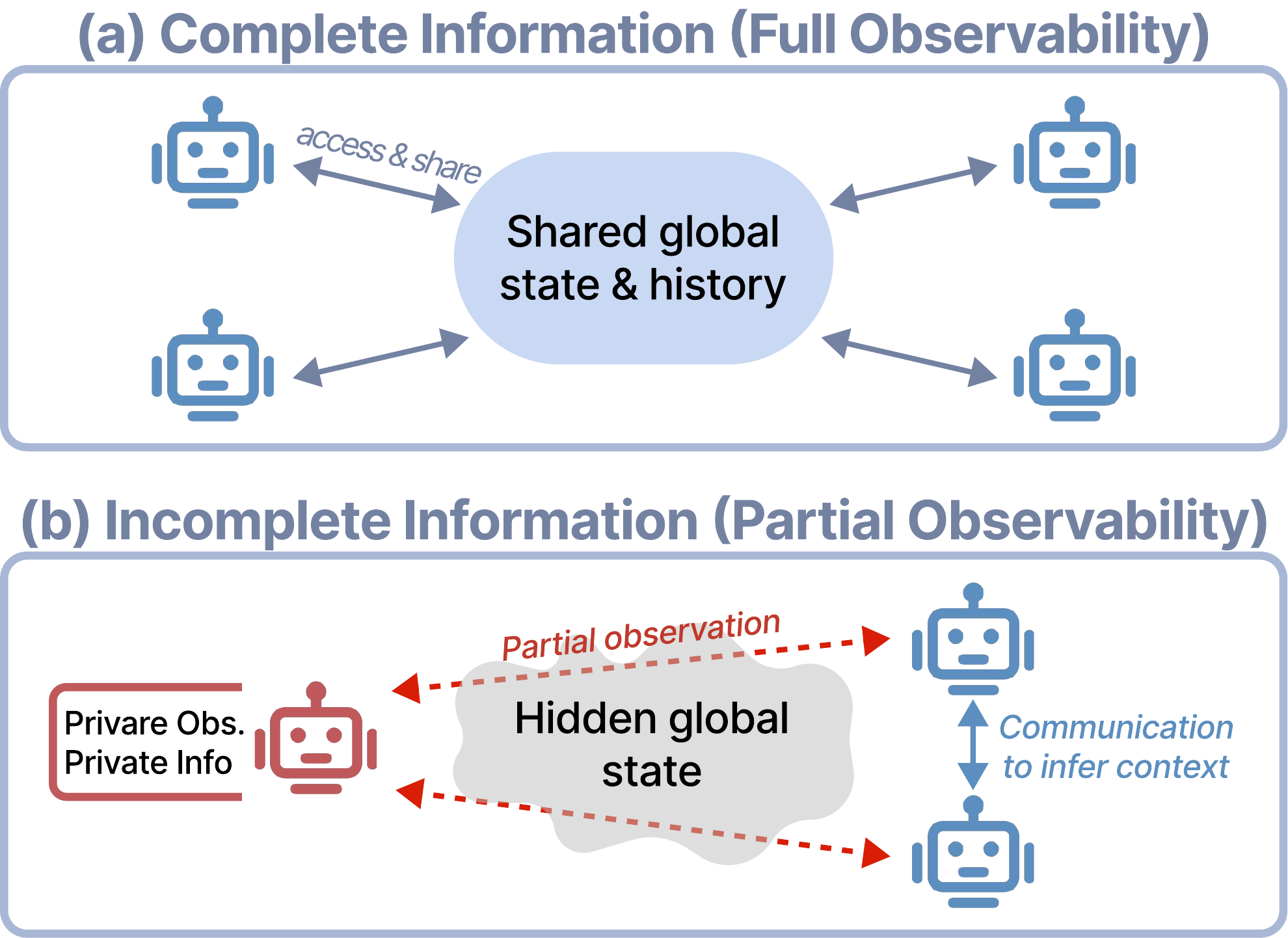}
    \vspace{-1mm}
    \caption{Illustration of information structures among LLM-based players.}
    \label{fig:info}
    \vspace{-4mm}
\end{figure}

The information structure of a system fundamentally determines the strategic complexity and coordination potential among agents.  
Let $\mathcal{I}_i(s)$ denote the information available to agent $i$ at state $s$.  
The strategic formulation of such systems can be expressed as a Bayesian game:
\[
U_i(a_i, a_{-i}) = \mathbb{E}_{\theta_i \sim p(\theta_i \mid \mathcal{I}_i)} \!\left[ R_i(s,a_i,a_{-i},\theta_i) \right],
\]
where $\theta_i$ represents private information or beliefs. This connects LLM-mediated interaction with classical incomplete-information games.  
The equilibrium efficiency thus depends critically on how $\mathcal{I}_i$ is structured and shared across agents.

\paragraph{Full Observability.}
Under complete information, all agents share the global state and can observe each other's actions and payoffs (as shown in Figure~\ref{fig:info} (a)).  
Such transparency enables joint planning but requires greater communication or computation.  
The AgentVerse framework~\cite{chen2023agentverse} demonstrates how full information sharing enhances performance in expert-agent collaboration.  
Similarly, the Chain-of-Agents~\cite{zhang2024chain} and COPPER~\cite{bo2024reflective} frameworks leverage complete observability to facilitate seamless sequential reasoning and reflective adaptation.
Full observability fosters emergent role-based collaboration and stronger generalization.
Even in simpler cooperative setups like CAMEL~\cite{li2023camel}, the shared full state empowers the two role-assigned LLMs to coordinate tightly and outperform single-agent baselines.
From a game-theoretic lens, complete information approximates a common-knowledge environment, where every player's beliefs about others' knowledge converge, simplifying equilibrium computation.

\paragraph{Partial Observability.}
By contrast, incomplete information is more realistic but challenging.
Each agent receives a signal or observation $\theta_i$, and must select actions according to a belief distribution $b_i(s) = P(s \mid \theta_i)$ (as shown in Figure~\ref{fig:info} (b)).  
The optimal policy in such environments maximizes the expected return:
\[
\pi_i^*(\theta_i) = \arg\max_{\pi_i} \mathbb{E}_{s \sim b_i,\, a_i \sim \pi_i(\theta_i)} \!\left[ R_i(s,a_i,a_{-i}) \right].
\]
Li et al.~\cite{MARL_2024} introduce a language-grounded multi-agent reinforcement learning (MARL) pipeline where agents learn to communicate in natural language, which accelerates the emergence of effective protocols and even generalizes zero-shot to new teammates.
Additional studies on reliable decision-making~\cite{reliable2024decision} examine how agents coordinate effectively under distributed or incomplete information. It demonstrates that simple voting or independent aggregation among agents often outperforms complex iterative feedback loops, as error propagation from multi-round dialogue may destabilize the system.

Incomplete information games have been explicitly tested, ranging from multi-issue negotiations~\cite{llmdelibera_2023} to games like Poker and auctions~\cite{GAMELLM_2024}. Both works reveal that while naive LLMs often struggle with large hidden states or deviate from game-theoretic rationality, their performance improves significantly with structured reasoning.
Benchmarks further illustrate the impact of limited information.
GTBench~\cite{duan2024gtbench} finds that LLMs fail completely informed deterministic games (like \texttt{Tic-Tac-Toe}) but can remain competitive in chance-driven or stochastic games. 
Real-world style simulations also highlight these effects. 
Park et al.~\cite{park2023generative} show that 25 generative agents can coordinate complex social tasks through local memory and iterative dialogue.
Piao et al.~\cite{agentsociety_2025} scale this to 10,000 agents, demonstrating how macro-phenomena like polarization emerge from decentralized exchanges. 
Together, they prove that agents can bridge partial observability by using inter-agent communication to infer global context from local interactions.
\section{Discussion}
In this section, we first review representative benchmarks in this field. 
Section~\ref{sec:case} then analyzes specific examples to show how game theory empowers the design of LLM agents. 
Finally, Section~\ref{sec:future} identifies critical gaps and proposes potential research directions for the field.

\subsection{Benchmark}
Benchmarks for LLM-based multi-agent systems are categorized into two primary groups focusing on fundamental capabilities and specialized domain applications respectively.
\paragraph{General-purpose Cognitive Benchmarks.} 
These benchmarks evaluate coordination and reasoning by decoupling core capabilities from domain knowledge.
For instance, GAIA requires agents to execute multi-step tasks, such as retrieving historical data to perform calculations, to test their ability to decompose problems autonomously. 

\paragraph{Domain-specific evaluation frameworks.}
These shift the focus toward functional utility within vertical sectors. Environments, such as software engineering repositories or clinical diagnostic simulators, require agents to possess deep ontological knowledge and specialized tool-calling proficiency.
Evaluation here is centered on the agent's adherence to professional protocols and its efficiency in executing collaborative workflows that mimic real-world industry demands.

\begin{table*}[t]
    \centering
    \small
    \caption{\textbf{A Taxonomy of Representative MAS Benchmarks.}
    Benchmarks are grouped by application domain, reflecting the shift from general-purpose coordination to domain-specialized multi-agent collaboration.}
    \label{tab:mas-benchmarks}
    \vspace{-2mm}
    \renewcommand{\arraystretch}{1.25}
    \setlength{\tabcolsep}{7pt}
    \begin{tabularx}{\textwidth}{l l X}
        \toprule
        \rowcolor{gray!12}
        \textbf{Category} & \textbf{Domain} & \textbf{Representative Benchmarks} \\
        \midrule

        \textbf{General} 
        & General / Mixed
        & MultiAgentBench \cite{multiagentbench_2025};
          GAIA \cite{mialon2024gaia};
          AgentVerse \cite{chen2023agentverse};
          Magentic-One \cite{DBLP:journals/corr/abs-2411-04468} \\
        \midrule

        \multirow{4}{*}{\makecell[c]{\textbf{Specialized}}}
        & Data Science
        & MLE-bench \cite{DBLP:conf/iclr/ChanCJASMSLMPMW25};
          DSBench \cite{DBLP:conf/iclr/JingHWYYM0DY25};
          DABstep \cite{DBLP:journals/corr/abs-2506-23719} \emph{(autonomous ML engineering)} \\
        \cmidrule(lr){2-3}

        & Software Engineering
        & SWE-bench \cite{DBLP:conf/iclr/JimenezYWYPPN24};
          rSDE-bench \cite{hu2024self} \emph{(repository-level reasoning)} \\
        \cmidrule(lr){2-3}

        & Finance
        & FinBen \cite{xie2024finben};
          AI-Trader \cite{fan2025ai};
          FinGAIA \cite{DBLP:journals/corr/abs-2507-17186} \emph{(role-specialized decision making)} \\
        \cmidrule(lr){2-3}

        & Planning / Robotics
        & TravelPlanner \cite{DBLP:conf/icml/Xie0CZLTX024};
          MAP-THOR \cite{nayak2024mapthor};
          PARTNR \cite{DBLP:conf/iclr/ChangCCCDHKKMPP25} \emph{(long-horizon coordination)} \\
        \bottomrule
    \end{tabularx}
    \vspace{-3mm}
\end{table*}

\subsection{Case Studies}
\label{sec:case}

\begin{figure}[h]
    \centering
    \includegraphics[width=0.75\linewidth]{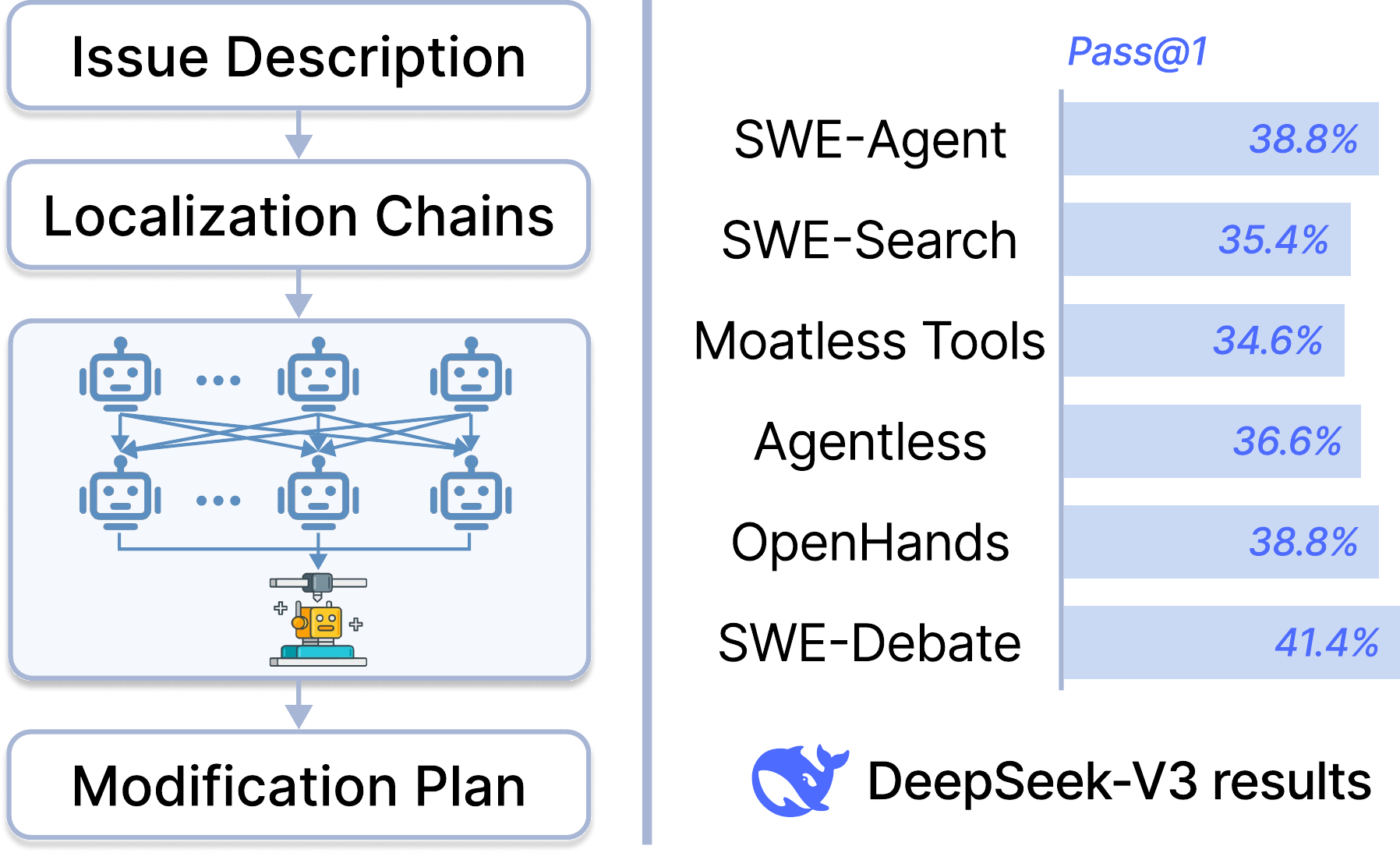}
    \vspace{-1mm}
    \caption{Framework and performance of SWE-Debate. (Left) The workflow incorporates a multi-agent debate mechanism to iteratively refine modification plans. (Right) This competitive architecture achieves a SOTA 41.4\% Pass@1 (DeepSeek-V3), significantly outperforming non-competitive baselines and validating the efficacy of adversarial interactions.}
    \vspace{-4mm}
    \label{fig:case1}
\end{figure}

\begin{purplebox}
\textbf{Case One:} A carefully orchestrated competitive mechanism in MAS is pivotal for automated software repair.
\end{purplebox}
For example, SWE-Debate~\cite{swedebate_2025} formulates software resolution as a non-cooperative game, leveraging multi-path fault traces as initial strategies to escape single-agent local optima. 
Through structured three-round debates, agents are forced into a ``defend-and-critique'' cycle that exposes latent architectural trade-offs. 
This adversarial tension allows the discriminator to synthesize robust evidence, resolving ambiguities to reach a globally optimal equilibrium.


\begin{figure}[h]
    \centering
    \includegraphics[width=0.99\linewidth]{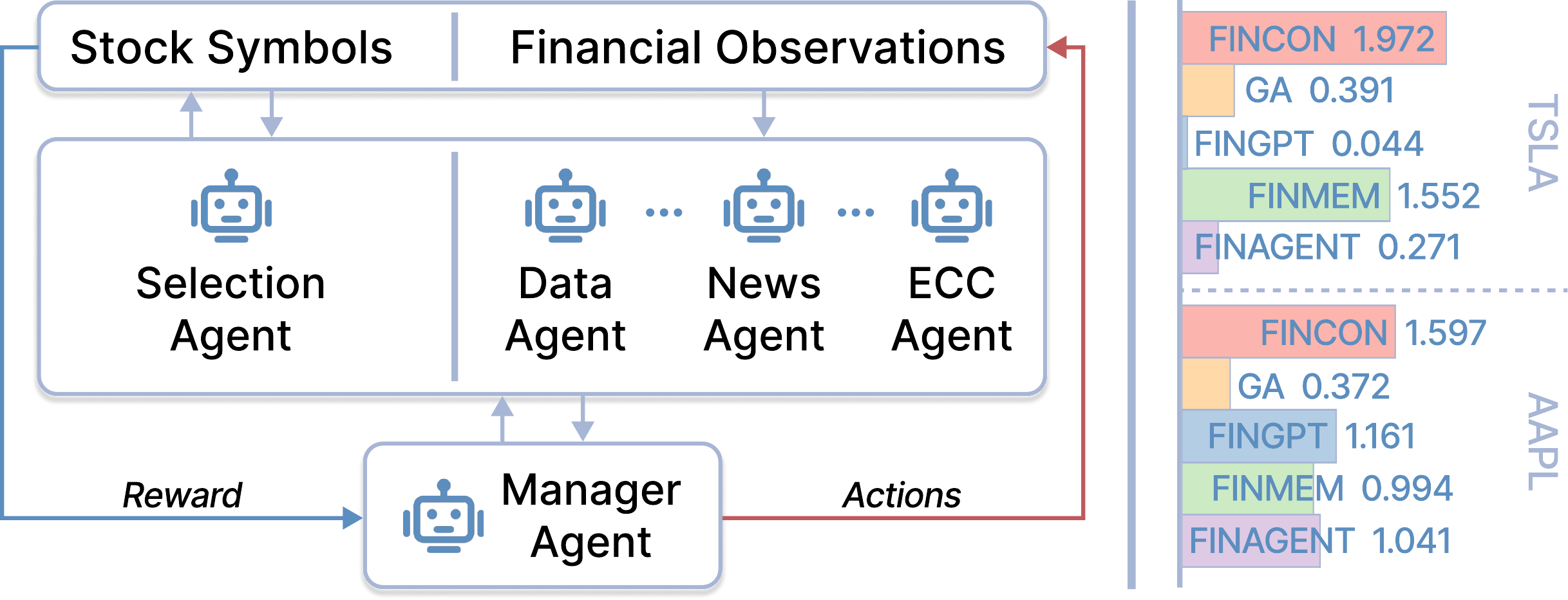}
    \vspace{-1mm}
    \caption{Reward shaping in the FinCon framework.
(Left) A decentralized multi-agent financial system coordinated by a manager agent, where agent behaviors are guided through shaped rewards derived from heterogeneous financial observations.
(Right) Empirical comparison illustrating how reward shaping and regulatory penalties influence agent performance across assets and backbone models.}
    \vspace{-4mm}
    \label{fig:case2}
\end{figure}

\begin{purplebox}
\textbf{Case Two:} Reward shaping and regulation can align decentralized incentives with systemic stability in financial decision-making.
\end{purplebox}
Financial markets are inherently complex systems characterized by intricate interdependencies, with theoretical foundations deeply rooted in market microstructure under asymmetric information~\cite{XuHTCLLZ25}. In such environments where agents face a fundamental conflict between private profit $v_{i}^{self}$ and market stability $v^{col}$, the FinCon~\cite{zhang2024fincon} framework demonstrate that modifying reward signals is essential for ensuring ethical and stable outcomes. 
From a game-theoretic perspective, by applying reward shaping through potential functions $\Phi_i(s)$, the system accelerates convergence toward cooperative goals without altering the optimal policy, while regulatory mechanisms integrate penalty functions $\Psi_i(s,a)$ to discourage undesirable behaviors like resource misuse or ethical violations. 
These structured reward interventions, which can include verbal feedback or self-penalization, effectively steer decentralized LLM agents toward a stable equilibrium where individual motives are aligned with the collective good.

\subsection{Future Directions}
\label{sec:future}
\paragraph{Hierarchical Superagent Orchestration.}
Current MAS largely rely on predefined protocols or static role assignments. Future research should transition toward Autonomous Superagents characterized by two core functionalities: 
\begin{enumerate}
    \item \textbf{Resource Planning}: The rational orchestration of existing agents and tools through high-level meta-coordination to resolve complex, long-context tasks.
    \item \textbf{Generative Agent Synthesis}: The dynamic creation of specialized novel agents tailored to emergent environmental requirements.
\end{enumerate}
These systems must act as mechanism designers to resolve architectural trade-offs via collaborative synthesis or structured adversarial tension.

\paragraph{Agentic Evolution.}
To transcend the limitations of fixed-policy interactions, Agentic RL serves as a critical paradigm for continuous strategy evolution.
Research should model prompt selection as strategic actions, identifying a ``reasoning equilibrium'' that accounts for LLMs' bounded rationality.
Leveraging competitive self-play and turn-level advantage estimation, agents can update parameters dynamically.
This iterative process enables convergence toward optimal strategies in both zero-sum and cooperative language games.

\paragraph{Theoretical Formalization}
Current solutions rely on heuristic implementations but lack strict mathematical formalization. This limits the theoretical analysis of multi-agent interactions. Future work should establish a rigorous framework to model decision-making under incomplete information, such as through the lens of Bayesian games. A key challenge involves designing incentive-compatible protocols that guarantee truthful signaling and align decentralized actions with systemic stability.

\vspace{-2mm}
\section{Conclusion}
We provide a comprehensive survey of LLM-based MAS through a game-theoretic lens, offering a unified theoretical foundation for this rapidly evolving field.
By organizing current research around the core elements of players, strategies, payoffs, and information, we establish a systematic framework to categorize diverse agent behaviors and interaction dynamics.
Our analysis reveals that while LLMs excel in high-level reasoning and strategic communication, significant gaps remain in robust equilibrium selection and incentive compatibility in complex, partially observable environments.
We conclude by highlighting forward-looking research directions, emphasizing that the integration of classical game theory with LLMs will be pivotal for developing more reliable, autonomous, and socially intelligent multi-agent systems.






\bibliographystyle{named}
\bibliography{ijcaishort26}

@article{nash1950equilibrium,
  author = "John Nash",
  title = "Equilibrium Points in N-Person Games",
  journal = "Proceedings of the National Academy of Sciences",
  volume = "36",
  number = "1",
  pages = "48--49",
  year = "1950"
}

@incollection{sandholm1999distributed,
  author = "Tuomas W. Sandholm",
  title = "Distributed Rational Decision Making",
  booktitle = "Multiagent Systems: A Modern Approach to Distributed Artificial Intelligence",
  publisher = "MIT Press",
  pages = "201--258",
  year = "1999"
}

@inproceedings{littman1994markov,
  author = "Michael L. Littman",
  title = "Markov Games as a Framework for Multi-Agent Reinforcement Learning",
  booktitle = {Proc. ICML},
  pages = "157--163",
  year = "1994"
}

@article{epstein1999agent,
  author = "Joshua M. Epstein",
  title = "Agent-Based Computational Models and Generative Social Science",
  journal = "Complexity",
  volume = "4",
  number = "5",
  pages = "41--60",
  year = "1999"
}

@inproceedings{wei2022chain,
  author = "Jason Wei and Xuezhi Wang and Dale Schuurmans and others",
  title = "Chain-of-Thought Prompting Elicits Reasoning in Large Language Models",
  booktitle = {Proc. NeurIPS},
  volume = "35",
  pages = "24824--24837",
  year = "2022"
}

@inproceedings{huang2022language,
  author = "Wenlong Huang and Pieter Abbeel and Deepak Pathak and Igor Mordatch",
  title = "Language Models as Zero-Shot Planners: Extracting Actionable Knowledge for Embodied Agents",
  booktitle = {Proc. ICML},
  pages = "9118--9147",
  year = "2022"
}

@article{park2023generative,
  author = "Joon Sung Park and Joseph C. O'Brien and Carrie J. Cai and others",
  title = "Generative Agents: Interactive Simulacra of Human Behavior",
  journal = {Proc. UIST},
  pages = "1--22",
  year = "2023"
}

@article{agentsociety_2025,
  title={Agentsociety: Large-scale simulation of llm-driven generative agents advances understanding of human behaviors and society},
  author={Piao, Jinghua and Yan, Yuwei and Zhang, Jun and others},
  journal={arXiv preprint arXiv:2502.08691},
  year={2025}
}

@article{li2023camel,
  author = "Guohao Li and Hasan Abed Al Kader Hammoud and Hani Itani and others",
  title = "CAMEL: Communicative Agents for Mind Exploration of Large Language Model Society",
  journal = {Proc. NeurIPS},
  volume = "36",
  pages = "51991--52008",
  year = "2023"
}

@inproceedings{
wu2023autogen,
title={AutoGen: Enabling Next-Gen {LLM} Applications via Multi-Agent Conversations},
author={Qingyun Wu and Gagan Bansal and Jieyu Zhang and others},
booktitle={First Conference on Language Modeling},
year={2024}
}

@inproceedings{
hong2023metagpt,
title={Meta{GPT}: Meta Programming for A Multi-Agent Collaborative Framework},
author={Sirui Hong and Mingchen Zhuge and Jonathan Chen and others},
booktitle={Proc. ICLR},
year={2024}
}

@inproceedings{
chen2023agentverse,
title={AgentVerse: Facilitating Multi-Agent Collaboration and Exploring Emergent Behaviors},
author={Weize Chen and Yusheng Su and Jingwei Zuo and others},
booktitle={Proc. ICLR},
year={2024}
}

@inproceedings{qian2024chatdev,
  title={Chatdev: Communicative agents for software development},
  author={Qian, Chen and Liu, Wei and Liu, Hongzhang and others},
  booktitle={Proc. ACL},
  pages={15174--15186},
  year={2024}
}

@article{swedebate_2025,
  title={Swe-debate: Competitive multi-agent debate for software issue resolution},
  author={Li, Han and Shi, Yuling and Lin, Shaoxin and others},
  journal={arXiv preprint arXiv:2507.23348},
  year={2025}
}

@inproceedings{agashe2025llmcoordinationevaluatinganalyzingmultiagent,
    title = "{LLM}-Coordination: Evaluating and Analyzing Multi-agent Coordination Abilities in Large Language Models",
    author = "Agashe, Saaket  and Fan, Yue  and Reyna, Anthony  and  Wang, Xin Eric",
    booktitle = {Proc. NAACL},
    year = "2025",
    pages = "8038--8057"
}

@article{ashery2025emergent,
  title={Emergent social conventions and collective bias in LLM populations},
  author={Ashery, Ariel Flint and Aiello, Luca Maria and Baronchelli, Andrea},
  journal={Sci. Adv.},
  volume={11},
  number={20},
  pages={eadu9368},
  year={2025}
}

@article{GAMELLM_2024,
  title={Game-theoretic llm: Agent workflow for negotiation games},
  author={Hua, Wenyue and Liu, Ollie and Li, Lingyao and others},
  journal={arXiv preprint arXiv:2411.05990},
  year={2024}
}

@article{MARSHAL_2025,
  title={MARSHAL: Incentivizing Multi-Agent Reasoning via Self-Play with Strategic LLMs},
  author={Yuan, Huining and Xu, Zelai and Tan, Zheyue and others},
  journal={arXiv preprint arXiv:2510.15414},
  year={2025}
}

@article{zhang2024chain,
  title={Chain of agents: Large language models collaborating on long-context tasks},
  author={Zhang, Yusen and Sun, Ruoxi and Chen, Yanfei and others},
  journal={Proc. NeurIPS},
  volume={37},
  pages={132208--132237},
  year={2024}
}

@article{bo2024reflective,
  title={Reflective multi-agent collaboration based on large language models},
  author={Bo, Xiaohe and Zhang, Zeyu and Dai, Quanyu and others},
  journal={Proc. NeurIPS},
  volume={37},
  pages={138595--138631},
  year={2024}
}

@inproceedings{duan2024gtbench,
  author = {Jinhao Duan and Renming Zhang and James Diffenderfer and others},
  title = {GTBench: Uncovering the Strategic Reasoning Limitations of LLMs via Game-Theoretic Evaluations},
  booktitle = {Proc. NeurIPS},
  year = {2024}
}

@inproceedings{tang2024explaining,
  title={Explaining decisions of agents in mixed-motive games},
  author={Orner, Maayan and Maksimov, Oleg and Kleinerman, Akiva and others},
  booktitle={Proc. AAAI},
  pages={23267--23275},
  year={2025}
}

@article{gradient2024risk,
  title={Risk analysis techniques for governed llm-based multi-agent systems},
  author={Reid, Alistair and O'Callaghan, Simon and Carroll, Liam and Caetano, Tiberio},
  journal={arXiv preprint arXiv:2508.05687},
  year={2025}
}

@article{reasoning_nash_2025,
  title={Reasoning and behavioral equilibria in LLM-Nash games: From mindsets to actions},
  author={Zhu, Quanyan},
  journal={arXiv preprint arXiv:2507.08208},
  year={2025}
}

@article{selfplay_survey_2024,
  title={A survey on self-play methods in reinforcement learning},
  author={Zhang, Ruize and Xu, Zelai and Ma, Chengdong and others},
  journal={arXiv preprint arXiv:2408.01072},
  year={2024}
}

@article{spiral_2025,
  title={SPIRAL: Self-Play on Zero-Sum Games Incentivizes Reasoning via Multi-Agent Multi-Turn Reinforcement Learning},
  author={Liu, Bo and Guertler, Leon and Yu, Simon and others},
  journal={arXiv preprint arXiv:2506.24119},
  year={2025}
}

@article{hu2024cory,
  title={Coevolving with the other you: Fine-tuning llm with sequential cooperative multi-agent reinforcement learning},
  author={Ma, Hao and Hu, Tianyi and Pu, Zhiqiang and others},
  journal={Proc. NeurIPS},
  volume={37},
  pages={15497--15525},
  year={2024}
}

@inproceedings{XuHTCLLZ25,
  author       = {Yuanjian Xu and
                  Jianing Hao and
                  Guang Zhang and others},
  title        = {FinRipple: Aligning Large Language Models with Financial Market for
                  Event Ripple Effect Awareness},
  booktitle    = {Proc. ACL Findings},
  year         = {2025}
}

@article{zhang2024fincon,
  title={Fincon: A synthesized llm multi-agent system with conceptual verbal reinforcement for enhanced financial decision making},
  author={Yu, Yangyang and Yao, Zhiyuan and Li, Haohang and others},
  journal={Proc. NeurIPS},
  volume={37},
  pages={137010--137045},
  year={2024}
}

@inproceedings{shen2024proagent,
  title={Proagent: building proactive cooperative agents with large language models},
  author={Zhang, Ceyao and Yang, Kaijie and Hu, Siyi and others},
  booktitle={Proc. AAAI},
  pages={17591--17599},
  year={2024}
}

@inproceedings{MARL_2024,
 author = {Li, Huao and Mahjoub, Hossein Nourkhiz and Chalaki, Behdad and others},
 booktitle = {Proc. NeurIPS},
 doi = {10.52202/079017-2790},
 pages = {87908--87933},
 title = {Language Grounded Multi-agent Reinforcement Learning with Human-interpretable Communication},
 volume = {37},
 year = {2024}
}

@article{reliable2024decision,
  title={Reliable Decision-Making for Multi-Agent LLM Systems},
  author={Lee, Xian Yeow and Akatsuka, Shunichi and Vidyaratne, Lasitha and others},
  journal={arXiv preprint arXiv:2406.04092},
  year={2025}
}

@article{llmdelibera_2023,
  title={LLM-deliberation: Evaluating LLMs with interactive multi-agent negotiation game},
  author={Abdelnabi, Sahar and Gomaa, Amr and Sivaprasad, Sarath and others},
  journal={arXiv preprint arXiv:2309.17234},
  year={2023}
}

@book{multiagent_2008,
  title={Multiagent systems: Algorithmic, game-theoretic, and logical foundations},
  author={Shoham, Yoav and Leyton-Brown, Kevin},
  year={2008},
  publisher={Cambridge University Press}
}

@inproceedings{multiagentbench_2025,
  title={{M}ulti{A}gent{B}ench : Evaluating the Collaboration and Competition of {LLM} agents},
  author={Zhu, Kunlun and Du, Hongyi and Hong, Zhaochen and others},
  booktitle={Proc. ACL},
  pages={8580--8622},
  year={2025}
}

@article{gamebench_2024,
  title={Gamebench: Evaluating strategic reasoning abilities of llm agents},
  author={Costarelli, Anthony and Allen, Mat and Hauksson, Roman and others},
  journal={arXiv preprint arXiv:2406.06613},
  year={2024}
}

@article{selfplaying_2024,
  title={Self-playing adversarial language game enhances llm reasoning},
  author={Cheng, Pengyu and Dai, Yong Recognition and Hu, Tianhao and others},
  journal={Proc. NeurIPS},
  volume={37},
  pages={126515--126543},
  year={2024}
}

@inproceedings{mechanism_design_2024,
  title={Mechanism design for large language models},
  author={Duetting, Paul and Mirrokni, Vahab and Paes Leme, Renato and others},
  booktitle={Proc. ACM Web Conf.},
  pages={144--155},
  year={2024}
}

@article{everyone_2025,
  title={Everyone Contributes! Incentivizing Strategic Cooperation in Multi-LLM Systems via Sequential Public Goods Games},
  author={Liang, Yunhao and Qu, Yuan and Yang, Jingyuan and others},
  journal={arXiv preprint arXiv:2508.02076},
  year={2025}
}

@article{kamenica2011bayesian,
  title={Bayesian persuasion},
  author={Kamenica, Emir and Gentzkow, Matthew},
  journal={American Economic Review},
  volume={101},
  number={6},
  pages={2590--2615},
  year={2011}
}

@book{fudenberg1998theory,
  title={The theory of learning in games},
  author={Fudenberg, Drew and Levine, David K},
  volume={2},
  year={1998},
  publisher={MIT press}
}

@article{meta2022human,
  title={Human-level play in the game of Diplomacy by combining language models with strategic reasoning},
  author={Meta Fundamental AI Research Diplomacy Team (FAIR)† and Bakhtin, Anton and Brown, Noam and others},
  journal={Science},
  volume={378},
  number={6624},
  pages={1067--1074},
  year={2022}
}

@article{silver2016alphago,
  title={Mastering the Game of Go with Deep Neural Networks and Tree Search},
  author={Silver, David and others},
  journal={Nature},
  volume={529},
  number={7587},
  pages={484--489},
  year={2016}
}

@book{mas1995microeconomic,
  title={Microeconomic theory},
  author={Mas-Colell, Andreu and Whinston, Michael Dennis and Green, Jerry R and others},
  volume={1},
  year={1995},
  publisher={Oxford university press New York}
}

@inproceedings{
mialon2024gaia,
title={{GAIA}: a benchmark for General {AI} Assistants},
author={Gr{\'e}goire Mialon and Cl{\'e}mentine Fourrier and Thomas Wolf and Yann LeCun and Thomas Scialom},
booktitle={Proc. ICLR},
year={2024}
}

@article{DBLP:journals/corr/abs-2411-04468,
  title={Magentic-One: {A} Generalist Multi-Agent System for Solving Complex Tasks},
   author = {Adam Fourney and Gagan Bansal and Hussein Mozannar and others},
  journal={arXiv preprint arXiv:2411.04468},
  year={2024}
}

@inproceedings{DBLP:conf/iclr/ChanCJASMSLMPMW25,
  author       = {Jun Shern Chan and
                  Neil Chowdhury and
                  Oliver Jaffe and others},
  title        = {MLE-bench: Evaluating Machine Learning Agents on Machine Learning
                  Engineering},
  booktitle    = {Proc. ICLR},
  year         = {2025}
}

@inproceedings{DBLP:conf/iclr/JingHWYYM0DY25,
  author       = {Liqiang Jing and
                  Zhehui Huang and
                  Xiaoyang Wang and others},
  title        = {DSBench: How Far Are Data Science Agents from Becoming Data Science Experts?},
  booktitle    = {Proc. ICLR},
  year         = {2025}
}

@article{DBLP:journals/corr/abs-2506-23719,
  title={DABstep: Data Agent Benchmark for Multi-step Reasoning},
  author={Alex Egg and Martin Iglesias Goyanes and Friso Kingma and others},
  journal={arXiv preprint arXiv:2506.23719},
  year={2025}
}

@inproceedings{DBLP:conf/iclr/JimenezYWYPPN24,
  author       = {Carlos E. Jimenez and
                  John Yang and
                  Alexander Wettig and others},
  title        = {SWE-bench: Can Language Models Resolve Real-world Github Issues?},
  booktitle    = {Proc. ICLR},
  year         = {2024}
}

@article{hu2024self,
  title={Self-evolving multi-agent collaboration networks for software development},
  author={Hu, Yue and Cai, Yuzhu and Du, Yaxin and others},
  journal={arXiv preprint arXiv:2410.16946},
  year={2024}
}

@article{xie2024finben,
  title={Finben: A holistic financial benchmark for large language models},
  author={Xie, Qianqian and Han, Weiguang and Chen, Zhengyu and others},
  journal={Proc. NeurIPS},
  volume={37},
  pages={95716--95743},
  year={2024}
}

@article{fan2025ai,
  title={AI-Trader: Benchmarking Autonomous Agents in Real-Time Financial Markets},
  author={Fan, Tianyu and Yang, Yuhao and Jiang, Yangqin and others},
  journal={arXiv preprint arXiv:2512.10971},
  year={2025}
}

@article{DBLP:journals/corr/abs-2507-17186,
  author = {Lingfeng Zeng and Fangqi Lou and Zixuan Wang and others},
  title = {FinGAIA: {A} Chinese Benchmark for {AI} Agents in Real-World Financial Domain},
  journal={arXiv preprint arXiv:2507.17186},
  year={2025}
}

@inproceedings{DBLP:conf/icml/Xie0CZLTX024,
  author       = {Jian Xie and Kai Zhang and Jiangjie Chen and others},
  title        = {TravelPlanner: {A} Benchmark for Real-World Planning with Language
                  Agents},
  booktitle    = {Proc. ICML},
  year         = {2024}
}

@inproceedings{nayak2024mapthor,
title={{MAP}-{THOR}: Benchmarking Long-Horizon Multi-Agent Planning Frameworks in Partially Observable Environments},
author={Siddharth Nayak and Adelmo Morrison Orozco and Marina Ten Have and others},
booktitle={Proc. Multi-modal Found. Model Meets Embodied AI Workshop, ICML 2024},
year={2024}
}

@inproceedings{DBLP:conf/iclr/ChangCCCDHKKMPP25,
  author       = {Matthew Chang and
                  Gunjan Chhablani and
                  Alexander Clegg and others},
  title        = {{PARTNR:} {A} Benchmark for Planning and Reasoning in Embodied Multi-agent Tasks},
  booktitle    = {Proc. ICLR},
  year         = {2025}
}

\end{document}